\documentclass{emulateapj}
\usepackage{apjfonts}
\usepackage{amsmath}

\def\mr#1{\mathrm{#1}}

\newcounter{ichi}
\newcounter{ni}
\newcounter{san}
\newcounter{yon}
\makeatletter
\newsavebox{\@parc@ption}
\def\parcaption#1{%
\sbox{\@parc@ption}{\shortstack[l]{#1}}%
>\setbox\@tempboxa\hbox{\csname fnum@\@captype\endcsname}%
\@tempdima\columnwidth \advance\@tempdima-\wd\@tempboxa
\@tempdimb\@tempdima %<-- maximum length is set here
\ifdim\wd\@parc@ption>\@tempdimb \@tempdima\@tempdimb
\else\@tempdima\wd\@parc@ption\fi
\sbox{\@tempboxa}{\parbox[t]{\@tempdima}{#1}}%
\caption{\usebox{\@tempboxa}}}
\makeatother

\slugcomment{}

\shorttitle{
Closure Relations for $e^{\pm}$ Pair-Signatures in GRBs
}
\shortauthors{Murase and Ioka}

\begin{document}

%% LaTeX will automatically break titles if they run longer than
%% one line. However, you may use \\ to force a line break if
%% you desire.

\title{
Closure Relations for $e^{\pm}$ Pair-Signatures in Gamma-Ray Bursts
%Pair-Signatures as a Diagnostic Tool\\
%of The Fireball in Gamma-Ray Bursts
}

%% Use \author, \affil, and the \and command to format
%% author and affiliation information.
%% Note that \email has replaced the old \authoremail command
%% from AASTeX v4.0. You can use \email to mark an email address
%% anywhere in the paper, not just in the front matter.
%% As in the title, use \\ to force line breaks.

\author{Kohta Murase\altaffilmark{1} and Kunihito Ioka\altaffilmark{2,3}}

%% Notice that each of these authors has alternate affiliations, which
%% are identified by the \altaffilmark after each name.  Specify alternate
%% affiliation information with \altaffiltext, with one command per each
%% affiliation.

\altaffiltext{1}{YITP, Kyoto University, Kyoto, 606-8502, Japan}
\altaffiltext{2}{Department of Physics, Kyoto University, Kyoto
606-8502, Japan}
\altaffiltext{3}{Theory Division, KEK (High Energy Accelerator 
Research Organization), 1-1 Oho, Tsukuba 305-0801, Japan}
%% Mark off your abstract in the ``abstract'' environment. In the manuscript
%% style, abstract will output a Received/Accepted line after the
%% title and affiliation information. No date will appear since the author
%% does not have this information. The dates will be filled in by the
%% editorial office after submission.

\begin{abstract}
We present recipes to diagnose
the fireball of gamma-ray bursts (GRBs) by combining
observations of $e^{\pm}$
pair-signatures (the pair-annihilation line and the cutoff
energy due to the pair-creation process).
Our recipes are largely model-independent and
extract information even from 
the non-detection of either pair-signature.
We evaluate physical quantities such as
the Lorentz factor, optical depth and pair-to-baryon ratio,
only from the observable quantities.
In particular, we can test whether the prompt emission of GRBs
comes from the pair/baryonic photosphere or not.
The future-coming Gamma-Ray Large Area Space Telescope
(GLAST) satellite will provide us with good chances to use our recipes
by detecting or non-detecting pair-signatures.
\end{abstract}

%% Keywords should appear after the \end{abstract} command. The uncommented
%% example has been keyed in ApJ style. See the instructions to authors
%% for the journal to which you are submitting your paper to determine
%% what keyword punctuation is appropriate.

%% Authors who wish to have the most important objects in their paper
%% linked in the electronic edition to a data center may do so in the
%% subject header.  Objects should be in the appropriate "individual"
%% headers (e.g. quasars: individual, stars: individual, etc.) with the
%% additional provision that the total number of headers, including each
%% individual object, not exceed six.  The \objectname{} macro, and its
%% alias \object{}, is used to mark each object.  The macro takes the object
%% name as its primary argument.  This name will appear in the paper
%% and serve as the link's anchor in the electronic edition if the name
%% is recognized by the data centers.  The macro also takes an optional
%% argument in parentheses in cases where the data center identification
%% differs from what is to be printed in the paper.


%% From the front matter, we move on to the body of the paper.
%% In the first two sections, notice the use of the natbib \citep
%% and \citet commands to identify citations.  The citations are
%% tied to the reference list via symbolic KEYs. The KEY corresponds
%% to the KEY in the \bibitem in the reference list below. We have
%% chosen the first three characters of the first author's name plus
%% the last two numeral of the year of publication as our KEY for
%% each reference.

\section{Introduction}
Gamma-ray burst (GRB) is one of the most mysterious objects in the universe. 
Various models are suggested, but no conclusive picture has been
obtained \citep[see reviews, e.g.,][]{Mes1,Zha1}. 
One of the leading models is the optically thin internal
shock model, where the prompt emission is explained by electromagnetic 
radiation from relativistic electrons accelerated in the internal
shocks \citep[see, e.g.,][]{Ree1}. 
One of the other leading models is the photospheric emission model, 
where the prompt emission comes from the photospheric radius $r_{\rm{ph}}$ at
which the Thomson optical depth is unity, i.e., $\tau =1$ 
\cite[see, e.g.,][]{Ree2}. The possibility 
that a fireball contains copious $e^{\pm}$ pairs 
(a pair-dominated fireball) is also discussed by many authors.
In particular, we recently proposed that
the pair photosphere is unstable and capable of making
the observed non-thermal spectrum with high radiative efficiency
\cite{Iok1}.
The existence of copious pairs can extend the photosphere 
compared to baryonic photosphere which is determined by 
baryon-related electrons. Such pairs could be produced via 
dissipation processes such as internal shocks and magnetic reconnection.  

Prompt gamma-rays are typically radiated at $\sim 100$
keV. Observationally, even more high-energy photons were detected by
the EGRET detector. Such high-energy emissions are theoretically
expected due to
radiation processes such as the synchrotron and/or inverse Compton
emission. Sufficiently high-energy photons cannot
avoid the pair-production process, which leads to the existence of the
cutoff energy due to pair-creation. On the other hand, there may be a
lot of pairs that can be seen as pair-annihilation lines via the 
pair-annihilation process \cite{Iok1,Pee1,Pee3}. Future-coming GLAST 
satellite is the suitable detector to observe such pair-signatures, 
a pair-annihilation line and/or cutoff energy.

Obviously, such pair-signatures (the pair-annihilation line and the cutoff 
energy due to the pair-creation process) have important information on 
the fireball of GRBs. For example, the cutoff energy due to
pair-creation has information on the bulk Lorentz factor of a
fireball. This possibility has already been investigated by several
authors \cite{Bar2,Lit1,Raz1}. However, there are few studies 
focusing on both of the pair-annihilation line and the cutoff
energy due to pair-creation. 

In this paper, we clarify that, combining both of pair-signatures,
we can get much information about the GRB fireball (\S~\ref{sec:test}).
Even if we can not detect either of pair-signatures,
the non-detection itself gives information (\S~\ref{sec:limit}).
We show that observations of pair-signatures allow us to evaluate 
the Lorentz factor, optical depth of a fireball and 
pair-to-baryon ratio and so on.
In particular, we derive these relations only from the observable quantities
and make discussions as model-independently as possible.
Our recipes are especially profitable to
test the pair photospheric emission model (\S~\ref{sec:implications}).

Throughout the paper, we shall assume that we know the gamma-ray spectrum 
in the wide energy range (e.g., the high-energy spectral index $\beta$
and so on), source redshift $z$ from other observations,
and hence the
luminosity $\varepsilon L_{\varepsilon}$ 
at given observed energy $\varepsilon$
from the observed flux (see Fig.~\ref{fig:spec}).
   
\section{Diagnosing the Fireball by $e^{\pm}$ Pair-Signatures}\label{sec:test}
\begin{figure}
\plotone{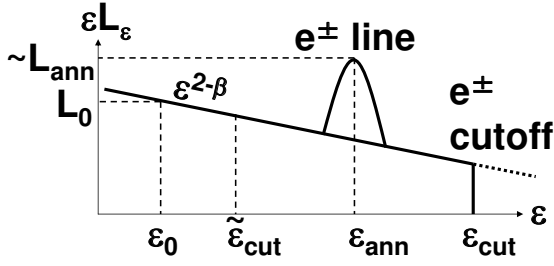}
\caption{
\footnotesize The schematic picture of the GRB spectrum
showing the observable quantities of pair-signatures
[$L_{\rm{ann}}$, $L_{0}$, $\varepsilon_0$,
$\varepsilon_{\rm{cut}}$, $\varepsilon_{\rm{ann}} \simeq \Gamma m_e c^2/(1+z)$
and $\beta$].
The recipes to constrain physical quantities of the GRB fireball only with
the observable quantities are the followings.
\hspace{15cm} \, 
(\Roman{ichi}) The case where we can observe both of the
pair-annihilation line and the cutoff energy due to
pair-creation, i.e., $\varepsilon_{\rm{ann}}$, $L_{\rm{ann}}$ and
${\varepsilon}_{\rm{cut}}$; 
If we also obtain the kinetic luminosity of baryons $L_p$,
we can measure $\tau$, $\tau_{\pm}$ and $n_{\pm}/n_p$
from Eqs.~(\ref{eq:tau2}), (\ref{eq:npmnp}) and (\ref{eq:taupm2}).
Without $L_p$, we obtain the inequalities~(\ref{eq:taupmtau}) from $L_p>0$,
while we have an upper limit on $\tau$ as well as lower limits on
$\tau_{\pm}$ and $n_{\pm}/n_p$
from the assumption $L_p \lesssim L_{\gamma}$ 
by replacing $L_p$ with $L_{\gamma}$ in 
Eqs.~(\ref{eq:tau2}), (\ref{eq:npmnp}) and (\ref{eq:taupm2}).
If the inequality (\ref{eq:pairdom}) is satisfied,
the fireball is pair-dominated, $\tau \approx \tau_{\pm}$,
and we can use Eqs.~(\ref{eq:taupm}) and (\ref{eq:Lpm0}) 
instead of Eqs.~(\ref{eq:tau2}) and (\ref{eq:Lpm2}).
\hspace{15cm} \,
(\Roman{ni}) The case where we only observe ${\varepsilon}_{\rm{cut}}$,
not $L_{\rm{ann}}$ and ${\varepsilon}_{\rm{ann}}$;
With $L_p$, we can give upper limits on $\tau$, $n_{\pm}/n_p$ and $\tau_{\pm}$
by replacing $L_{\rm{ann}}$ with
$L_{0} (\Gamma m_e c^2/(1+z) \varepsilon_{0})^{2-\beta}$
in Eqs.~(\ref{eq:tau2}), (\ref{eq:npmnp}) and (\ref{eq:taupm2}).
Without $L_p$, we obtain the inequality~(\ref{eq:taupm<1}) from $L_p>0$,
while we have an upper limit on $\tau$ from $L_p \lesssim L_{\gamma}$
by replacing $L_p$ and $L_{\rm{ann}}$ 
with $L_{\gamma}$ and $L_{0} (\Gamma m_e c^2/(1+z) \varepsilon_{0})^{2-\beta}$,
respectively, in Eq.~(\ref{eq:tau2}).
$\Gamma \, (<(1+z)\varepsilon_{\rm{cut}}/m_e c^2)$ 
should be acquired by other means.
\hspace{15cm} \,
(\Roman{san}) The case where we only observe $L_{\rm{ann}}$ and 
$\varepsilon_{\rm{ann}}$, not ${\varepsilon}_{\rm{cut}}$; 
We regard the observed maximum energy ${\varepsilon}_{\rm{max}}$
as the lower limit on the true cutoff energy 
${\varepsilon}_{\rm{cut}}$. 
With $L_p$, we can give upper limits on $\tau$ and $\tau_{\pm}$
as well as a lower limit on $n_{\pm}/n_p$
by replacing $f_{\rm{cut}}$ with
$f_{\max} \equiv f({\varepsilon}_{\rm{max}},\Gamma)$
in Eqs.~(\ref{eq:tau2}), (\ref{eq:npmnp}) and (\ref{eq:taupm2}).
Without $L_p$, we obtain the inequality~(\ref{eq:taupm<2}) from $L_p>0$,
while we obtain an upper limit on $\tau$ 
as well as a lower limit on $n_{\pm}/n_p$ from $L_p \lesssim L_{\gamma}$
by replacing $L_p$ and $f_{\rm{cut}}$ with
$L_{\gamma}$ and $f_{\max}$, respectively,
in Eqs.~(\ref{eq:tau2}) and (\ref{eq:npmnp}).
Such arguments can be also applied to the completely thin fireballs.
\hspace{15cm} \,
(\Roman{yon}) The recipes (\Roman{ichi})-(\Roman{san})
are especially valuable to test the pair photospheric emission model.
The inequalities (\ref{eq:taupmtau}), (\ref{eq:taupm<1}) and (\ref{eq:taupm<2})
are useful to constrain $\tau_{\pm}$.
This model gives $\tau_{\pm} \sim 1$
in the case (\Roman{ichi}), and
Eq.~(\ref{eq:ecuteann}) if $L_{\rm{ann}}$ is comparable to
the underlying continuum emission.
The photospheric radius can be also estimated. 
%from Eq.~(\ref{eq:rpm}). 
}
\label{fig:spec}
\end{figure}

Let us assume that we can find a pair-annihilation line in
the spectrum of the prompt emission (Fig.~\ref{fig:spec}), which
typically peaks at
\begin{equation}
\varepsilon_{\rm{ann}} \simeq \frac{\Gamma}{1+z} m_e c^2.
\end{equation}
The above expression is valid as long as pairs forming a
pair-annihilation line are non-relativistic. 
This is a reasonable assumption because
the cooling time of sufficiently relativistic pairs $t_{\rm{cool}}$ due to 
the magnetic and/or photon fields is usually much shorter than the 
pair-annihilation time $t_{\rm{ann}}$. However, we have to note that
the line would be 
%significantly 
broadened by dispersion of the Doppler
factor. Therefore, gamma-rays due to pair-annihilation will be
observed as a ``bump'' rather than a ``line''. There are several
possible reasons that make line-broadening. First, the order-of-unity 
distribution of the Lorentz factor in the emission region can make the
line broadened by order-of-unity even when pairs are
non-relativistic in the comoving frame. Second, the order-of-unity
line-broadening is also caused by the fact that we observe a section of the 
emission region with the opening angle $\sim 1/\Gamma$ rather than a
small spot. The Doppler factor towards the observer
is different by order-of-unity between the center and the edge 
of the observed emission region.
Third, the order-of-unity variation of the Lorentz
factor may also occur within the dynamical time. The recent obserbations 
may suggest the emission is radiatively very efficient \cite{ioka06,Zha2}. 
The efficient internal dissipation may make the fireball radiation-dominated. 
If so, the Lorentz factor increases as $\Gamma \propto r$, and the Lorentz 
factor varies by order-of-unity within the dynamical time.
%The Lorentz factor of 
%the outflow is likely to vary rapidly in the internal shock model, where 
%the Lorentz factor after an internal collision is about $\Gamma \sim 
%\sqrt{\Gamma_{f} \Gamma_{s}}$, in which $\Gamma_{f}$ and $\Gamma_{s}$ 
%are Lorentz factors of a rapid subshell and slow subshell, respectively.
%This order-of-unity variation of the Lorentz factor can also make 
%line-broadening by a factor of several. 
Therefore, we can expect that
all the three effects broaden the line by order-of-unity.

The total luminosity of the pair-annihilation line $L_{\rm{ann}}$
\cite{Cop1,Sve1},
the kinetic luminosity of pairs $L_{\pm}$
and the kinetic luminosity of baryons $L_p$
are given by 
\begin{eqnarray}
L_{\rm{ann}} &\simeq& \frac{3}{8}n_{+} n_{-} \sigma_{T}c 
(2 m_e c^2) (4 \pi r^2 \Delta^{\prime}) \Gamma^2, 
\label{eq:Lann}
\\ 
L_{\pm} &=& n_{\pm} c (2 m_e c^2) (4 \pi r^2 ) \Gamma^2, 
\label{eq:Lpm}
\\
L_p &=& n_p c (m_p c^2) (4 \pi r^2) \Gamma^2, 
\label{eq:Lp}
\end{eqnarray}
respectively.
Here $r$ is the emission radius, $\Delta^{\prime}$ is the 
comoving width of the emission region, and 
$n_{+}=n_{\pm}$, $n_{-}=n_{\pm}+n_{p}$, $n_{\pm}$ and $n_{p}$
are the comoving density of positrons, electrons,
$e^{\pm}$ pairs and baryon-related electrons, respectively.
We have assumed that most of the sufficiently relativistic 
pairs cool down in the dynamical time.
Combining expressions of $L_{\rm{ann}}$ and $L_{\pm}$ leads to
\begin{equation}
L_{\rm{ann}} \simeq \frac{3}{16} L_{\pm} \tau_{\pm}
\left(1+\frac{n_p}{n_{\pm}}\right),
\label{eq:LLt}
\end{equation}
where $\tau_{\pm} \simeq 2 n_{\pm} \sigma_T \Delta^{\prime}$
denotes the optical depth against pairs.

Pair-creation processes such as $\gamma \gamma \rightarrow e^{+} e^{-}$ and 
$e \gamma \rightarrow e e^{+} e^{-}$ prevent sufficiently high-energy
photons from escaping the source. Usually, the most important
pair-creation process is $\gamma \gamma \rightarrow e^{+} e^{-}$ \cite{Raz1}. 
The optical depth for this process $\tau_{\gamma \gamma}$
at some energy $\varepsilon$ can be evaluated for a given photon
spectrum. The elaborate evaluation of $\tau_{\gamma \gamma}$ is
possible if we know the spectrum in detail \citep[see,
e.g.,][]{Cop1,Bar1,Bar2,Gup1}. Here, we shall assume a power-law 
photon spectrum for simplicity, i.e., with the luminosity 
$\varepsilon L_{\varepsilon}(\varepsilon)
= L_0 (\varepsilon/\varepsilon_0)^{2-\beta}$
for $\beta>2$. Then, we have \cite{Gou1,Lig1,Sve2,Lit1,Bar1}
\begin{equation}
\tau_{\gamma \gamma}(\varepsilon) 
\simeq \xi (\beta) n_{\gamma} (\varepsilon_{\gamma} > \tilde{\varepsilon}) 
%\simeq \xi (\beta) \tilde{\varepsilon} n_{\varepsilon} (\tilde{\varepsilon})
\sigma_T \Delta^{\prime},
\quad \left[\tilde{\varepsilon} = 
\frac{(\Gamma m_e c^2)^2}{(1+z)^2 \varepsilon}\right],
\label{eq:taugg}
\end{equation}
where the comoving density of photons 
%with $\tilde{\varepsilon}$ is
whose energies are larger than $\tilde{\varepsilon}$ is
given by 
\begin{eqnarray}
n_{\gamma} (\varepsilon_{\gamma} > \tilde{\varepsilon})
%\tilde{\varepsilon} n_{\varepsilon} (\tilde{\varepsilon})
= \frac{L_{0}}{4 \pi r^2 \Gamma c \varepsilon_{0}(1+z)} 
\int_{\tilde{\varepsilon}} \frac{d \varepsilon_{\gamma}}{\varepsilon_{0}} 
{\left(\frac{\varepsilon_{\gamma}}{\varepsilon_{0}} \right)}^{-\beta}, 
%{\left(\frac{\tilde{\varepsilon}}{\varepsilon_{0}} \right)}^{-\beta+1}, 
\label{eq:ng}
\end{eqnarray}
and $\tilde{\varepsilon}$ is the
energy of a photon which interacts with the photon 
of energy $\varepsilon$ at the pair-creation threshold.
$\xi(\beta)$ is the numerical factor which depends on the photon
index \cite{Gou1,Lig1,Sve2,Cop1,Lit1,Bar1,Gup1}. $\xi(\beta)/(\beta-1)$ 
decreases with $\beta$, and its values are $\xi(\beta) = 11/90 \simeq 0.12$
and $\xi(\beta) = 7/75 \simeq 0.093$ 
for $\beta=2$ and $3$, respectively\footnote{When we assume
the isotropic photon spectrum with an infinite power-law, we have 
$\xi(\beta=2) = 11/90 \simeq 0.12$. This value is obtained by various 
authors, e.g., Gould \& Schr\'eder (1967), Svensson (1987), Baring (2006) and Gupta \& Zhang (2007). Lithwick \& Sari (2001) adopt a factor two smaller value
as is pointed in Zhang \& M\'esz\'aros (2001).}. For the isotropic photon
distribution with an infinite power-law, we can use $\xi(\beta)
\simeq 7(\beta-1)/(6{\beta}^{5/3}(\beta+1))$ for $1<\beta<7$
\cite{Sve2,Bar1}. 
Note that $L_0$ is related to the observed (time-resolved) flux $\varepsilon
F_{\varepsilon}(\varepsilon)$ by
\begin{equation}
\varepsilon F_{\varepsilon}(\varepsilon=\varepsilon_0) = 
\frac{L_0}{4 \pi d_L^2}, 
\end{equation}
where $d_L$ is the luminosity distance to the source.
Unless a fireball is completely thin, where all the photons 
can escape without attenuation, the cutoff energy 
$\varepsilon_{\rm{cut}}$ exists due to the pair-creation process 
$\gamma \gamma \rightarrow e^{+} e^{-}$, where $\tau_{\gamma
\gamma} (\varepsilon_{\rm{cut}})=1$ (Fig.~\ref{fig:spec}). 
With Eq.~(\ref{eq:Lpm}),
$\tau_{\gamma \gamma} (\varepsilon_{\rm{cut}})=1$ is rewritten as
\begin{equation}
1 = \tau_{\gamma \gamma} (\varepsilon_{\rm{cut}})
\simeq
\frac{L_{0}}{L_{\pm}} \tau_{\pm} f (\varepsilon_{\rm{cut}},\Gamma),
\label{eq:tau1}
\end{equation}
where
\begin{equation}
f (\varepsilon_{\rm{cut}},\Gamma) \simeq \xi (\beta)
\frac{\Gamma m_e c^2}{(1+z)\varepsilon_{0}}
\int_{\tilde{\varepsilon}_{\rm{cut}}} 
\frac{d \varepsilon_{\gamma}}{\varepsilon_{0}} 
{\left(\frac{\varepsilon_{\gamma}}{\varepsilon_{0}}\right)}^{-\beta}. 
%{\left(\frac{\tilde{\varepsilon}}{\varepsilon_{0}}\right)}^{-\beta+1}. 
\label{eq:fdef}
\end{equation}
Note that we may arbitrarily take $\varepsilon_0$ by adjusting $L_0$.
We also note that $\varepsilon_{\rm{cut}}$ is larger than 
$\varepsilon_{\rm{ann}}$, as long as $\varepsilon_{\rm{cut}}$ is
determined by the pair-creation process $\gamma \gamma \rightarrow e^{+} e^{-}$
 (and we have also assumed that electrons and positrons are accelerated enough 
to emit high-energy photons with $\varepsilon > \varepsilon_{\rm{cut}}$ via 
e.g., synchrotron or inverse Compton radiation processes). 
This is because an assumed photon spectrum has $\beta>1$ (which is
typically expected for prompt emissions), hence the photon number
density decreases with photon energies. Therefore, photons with
$\varepsilon \lesssim \varepsilon_{\rm{ann}}$ do not have enough target
photons with $\tilde{\varepsilon} \gtrsim \varepsilon_{\rm{ann}}$ in order
to be attenuated at $\varepsilon_{\rm{cut}} \lesssim
\varepsilon_{\rm{ann}}$. Otherwise, the created pairs would make the optical 
depth $\tau$ larger than unity. In this case, the cutoff energy is 
determined by the Compton down-scattering process rather than the pair-creation
process for the assumed spectrum (see, e.g., Lithwick \& Sari 2001).
Although we will hereafter focus on cases where $\varepsilon_{\rm{cut}}$ is 
determined by the pair-creation cutoff, there are possibilities of 
$\varepsilon_{\rm{cut}} \lesssim 
\varepsilon_{\rm{ann}}$ for $\tau \gtrsim 1$. We may be able to check $\varepsilon_{\rm{cut}}
 \lesssim \varepsilon_{\rm{ann}}$ and $\tau \gtrsim 1$, if we can observe the Lorentz 
factor $\Gamma$ by other means as well as the cutoff energy $\varepsilon_{\rm{cut}}$.
We would expect that high-energy gamma rays come from the region where $\tau \sim 1$, 
as long as the dissipation continues until $r \sim r_{\rm{ph}}$ and 
the emission from $r \sim r_{\rm{ph}}$ is not negligible.
This is because high-energy 
gamma rays from the region where $\tau \gg 1$ are significantly 
down-scattered. 
We would also expect that 
the GRB radiative efficiency is small (contrary to the observations)
if the prompt emission comes only from $\tau \gg 1$ 
since almost all energy goes into the afterglow.
We also note that in some models like the slow dissipation scenario 
\cite{Ghi1} high-energy photons with $\varepsilon >
\varepsilon_{\rm{cut}}$ 
may not be produced because electrons and positrons are not
accelerated enough \cite{Pee3}. 
  
%If the source cutoff energy is determined by
%another process, $\varepsilon_{\rm{cut}} \lesssim
%\varepsilon_{\rm{ann}}$ might be possible (e.g., when $\tau \gtrsim
%1$) \cite{Lit1}. 

\subsection{Closure Relations for the Pair-Dominated Fireball}
Now, let us assume the pair-dominated fireball,
$n_p < 2 n_{\pm}$, in this subsection.
Then we can solve Eqs.~(\ref{eq:LLt}) and (\ref{eq:tau1})
for the two unknown quantities
$\tau_{\pm}$ and $L_{\pm}$ as
\begin{eqnarray}
\tau \approx \tau_{\pm} 
&\simeq& \left(\frac{16}{3}
\frac{L_{\rm{ann}}}{L_{0} f(\varepsilon_{\rm{cut}},\Gamma)}
\right)^{1/2}
\label{eq:taupm}
\\
L_{\pm} &\simeq& 
\left(\frac{16}{3} L_0 L_{\rm{ann}}
f(\varepsilon_{\rm{cut}},\Gamma)
\right)^{1/2}.
\label{eq:Lpm0}
\end{eqnarray}
Remarkably, the above two quantities are expressed only by 
the observable quantities
[$L_{\rm{ann}}$, $L_{0}$, $\varepsilon_0$,
$\varepsilon_{\rm{cut}}$, $\varepsilon_{\rm{ann}} \simeq \Gamma m_e c^2/(1+z)$
and $\beta$], so that we can evaluate $\tau_{\pm} \approx \tau$ 
and $L_{\pm}$. Note that we have not assumed the frequently used
relation $r \approx 2 \Gamma \Delta'$, which is expected in the internal
shock model. Because we have not specified the model, our recipes are
largely model-independent in that sense. 
%Here, $t_{\rm{dyn}}$ is the dynamical time scale, which
%is roughly $\sim \Delta'/c$.
The absence of $\sigma_T$ in Eqs. (\ref{eq:taupm}) and (\ref{eq:Lpm0}) 
just comes from the fact that the pair-annihilation, pair-creation and Compton
scattering are all basic two-body interaction processes with
cross section $\sim \sigma_T$. 
Ambiguities arising from the transformation
between the comoving frame and observer frame are canceled,
because the transformation between the two frames is the same for 
$L_{0}$, $L_{\rm{\pm}}$ and $L_{\rm{ann}}$. 

Eqs. (\ref{eq:taupm}) and (\ref{eq:Lpm0}) are useful because they
enable us to estimate $\tau \approx \tau_{\pm}$ and $L_{\pm}$ from
observational quantities only, although there will be possible
uncertainties due to, e.g., observational difficulties in evaluation of 
$\varepsilon_{\rm{cut}}$, $\varepsilon_{\rm{cut}}$ and $L_{\rm{ann}}$. 
%and due to theoretical ambiguities in the relations. 
In Figs. 2 and 3, we demonstrate that we can obtain information on 
$\tau$ for a given burst (especially a given pulse). Observations of 
pair-signatures will enable us to plot the point in such a figure and 
to compare it with lines expressing optical depths. 
Of course, a line for a given $\tau$ is different among bursts with different 
parameter sets. However, we could see the tendency of the distribution of the optical 
depth for some bursts (or pulses) with a similar parameter set. In this case, 
lines for a given optical depth can be expressed as "a band" with a finite width.
We think that the plot without lines for optical depths may be also useful. 
More and more observations of pair-signatures will allow us to plot points with 
optical depths in the $\varepsilon_{\rm{cut}}-L_{\rm{ann}}$ plane.
  
The assumption $n_p < 2 n_{\pm}$ can be checked
posteriorly by the observations.
From Eqs.~(\ref{eq:Lpm}) and (\ref{eq:Lp}),
we have the condition for the fireball to be pair-dominated,
\begin{equation}
\frac{m_p L_{\pm}}{m_e L_p} \simeq \frac{2 n_{\pm}}{n_p} > 1,
\label{eq:pairdom}
\end{equation} 
which may be checked if we can measure $L_p$ from other observations.
For example, we could obtain
$L_p \sim L_p^{\rm{AG}}$, where $L_{p}^{\rm{AG}}$ is the kinetic
luminosity of baryons estimated from the afterglow observations.
Note that the inequality (\ref{eq:pairdom}) just means 
that the pair photospheric radius should be larger than the baryonic
photospheric radius, i.e., $r_{\rm{ph},\pm} > r_{{\rm{ph}}, p}$.
Especially, we have a closure relation $\tau_{\pm} \simeq 1$
for prompt emissions coming from a pair photosphere.

The kinetic luminosity of baryons may be usually less than the observed
gamma-ray luminosity, $L_{p} \lesssim L_{\gamma}$,
as inferred by recent observations that the prompt emission is radiatively
very efficient \cite{ioka06,Zha2}. It is not very convincing yet since 
we cannot measure the precise GRB energy at present. But once it is 
observationally established, we obtain the useful sufficient
condition. If the sufficient condition, $m_p L_{\pm}/ m_e L_{\gamma} > 1$, 
is satisfied, we can justify the pair-dominance in the 
inequality~(\ref{eq:pairdom}) by observations. This sufficient condition 
will be useful as we do not need to evaluate $L_p$.

\begin{figure}[bt]
\plotone{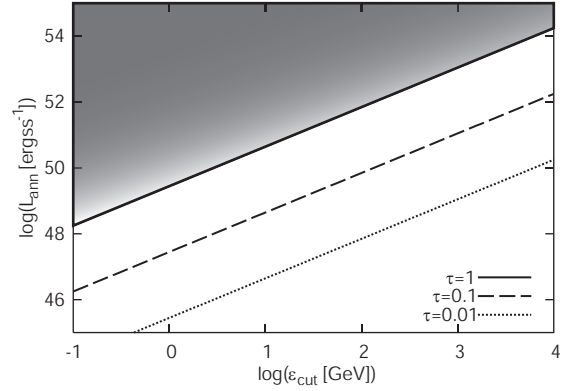}
\caption{
\footnotesize The relation between the cutoff energy
$\varepsilon_{\rm{cut}}$ and total luminosity of a pair annihilation
line $L_{\rm{ann}}$ for given optical depths $\tau$. Lines are
calculated by exploiting Eqs. (\ref{eq:fdef}) and (\ref{eq:tau2}). 
Used parameters are $L_{p} = {10}^{50}$ ergs $\rm{s}^{-1}$, 
$L_0 = {10}^{51}$ ergs $\rm{s}^{-1}$, $\varepsilon_{0} = {10}^{2.5}$ keV, 
$\Gamma = {10}^{2.5}$, $\beta=2.2$ and $z=0.1$. In this case, the
fireball is pair-dominated on this figure, and we can use Eq. 
(\ref{eq:taupm}) instead of Eq. (\ref{eq:tau2}). The shaded region
expresses $\tau \gtrsim 1$, where photons suffer from Compton
scatterings. If we can obtain necessary quantities such as
$L_{\rm{ann}}$ in Fig.~\ref{fig:spec}, we can estimate $\tau$ by
plotting observational quantities in this figure.
Note that $\varepsilon_{\rm{cut}}$ should be larger than
$\varepsilon_{\rm{ann}} \simeq \Gamma m_e c^2/(1+z)$ for typical
photon spectra as long as $\varepsilon_{\rm{cut}}$ is determined by
the pair-creation process. 
%(hence the region where $\varepsilon_{\rm{cut}}\lesssim \varepsilon_{\rm{ann}}$ is forbidden).
}
\label{fig:spec2}
\end{figure}

\subsection{More General Relations}
As shown in previous subsections, signatures of pair-annihilation 
and -creation are useful as a diagnostic tool of the pair-dominated 
fireball in GRBs. However, the fireball could not be pair-dominated, 
where the inequality (\ref{eq:pairdom}) is not satisfied. 
Taking into account of
the term $n_p/n_{\pm} \simeq 2 m_e L_p/m_p L_{\pm}$ in Eq.~(\ref{eq:LLt}),
we can derive the quadratic equation
for $L_{\pm}$ from Eqs.~(\ref{eq:LLt}) and (\ref{eq:tau1}),
and generalize Eqs.~(\ref{eq:taupm}) and (\ref{eq:Lpm0}) as
\begin{eqnarray}
\tau &\simeq& \left(\frac{16}{3}\frac{L_{\rm{ann}}}{L_0 f_{\rm{cut}}}
+\frac{m_e^2 L_p^2}{m_p^2 L_0^2 f_{\rm{cut}}^2} \right)^{1/2},
\label{eq:tau2}
\\
L_{\pm} &\simeq& 
\left(\frac{16}{3} L_0 L_{\rm{ann}} f_{\rm{cut}}
+\frac{m_e^2}{m_p^2} L_p^2 \right)^{1/2}
-\frac{m_e}{m_p} L_p,
\label{eq:Lpm2} 
\end{eqnarray}
where we have defined $f_{\rm{cut}} \equiv f(\varepsilon_{\rm{cut}},\Gamma)$,
and $\tau \simeq (2 n_{\pm} + n_p ) \sigma_T \Delta^{\prime} =
\tau_{\pm} (1+n_p/2n_{\pm})$ is the optical depth of the emission region. 
We can also evaluate the pair-to-baryon ratio 
and the optical depth against pairs as
\begin{eqnarray}
\frac{2n_{\pm}}{n_p}
&\simeq&\left(1+\frac{16 m_p^2 L_{\rm{ann}} L_0 
f_{\rm{cut}}}{3 m_e^2 L_p^2}\right)^{1/2}-1,
\label{eq:npmnp}
\\
\tau_{\pm}&\simeq&\frac{m_e L_p}{m_p L_0 f_{\rm{cut}}}
\left[\left(1+\frac{16 m_p^2 L_{\rm{ann}} L_0 
f_{\rm{cut}}}{3 m_e^2 L_p^2}\right)^{1/2}-1\right].
\label{eq:taupm2}
\end{eqnarray}
Compared to Eqs. (\ref{eq:taupm}) and (\ref{eq:Lpm0}), 
we additionally need information on the amount of baryons $L_p$ 
to obtain $\tau$, $L_{\pm}$ and $\tau_{\pm}$. 
If we take the no-pair limit $2 n_{\pm} \ll n_p$
in Eqs.~(\ref{eq:tau2}), (\ref{eq:Lpm2}) and (\ref{eq:taupm2}), 
we find that $\tau$ does not depend on
$L_{\rm{ann}}$, and $L_{\rm{\pm}}, \tau_{\pm} \rightarrow 0$, as expected.  

Even if we cannot estimate $L_p$, we have useful constraints 
only from pair-signatures. First, we can show 
\begin{equation}
\tau_{\pm} < 
\left(\frac{16}{3} \frac{L_{\rm{ann}}}{L_{0} f_{\rm{cut}}}
\right)^{1/2}
< \tau.
\label{eq:taupmtau}
\end{equation}
The above inequalities can be derived by exploiting $L_p > 0$ for 
Eqs.~(\ref{eq:tau2}) and (\ref{eq:taupm2}), respectively. 
Therefore, observations of pair-signatures give us the upper limit
on the optical depth against pairs. Especially, we can exclude 
the pair photospheric emission model when we have
$\tau_{\pm} \ll 1$. Second, with $L_{p} \lesssim L_{\gamma}$,
we can observationally give an upper limit on $\tau$ as well as
lower limits on $\tau_{\pm}$ and $2n_{\pm}/n_p$
by replacing $L_p$ with $L_{\gamma}$
in Eqs.~(\ref{eq:tau2}), (\ref{eq:npmnp}) and (\ref{eq:taupm2}).

\begin{figure}[bt]
\plotone{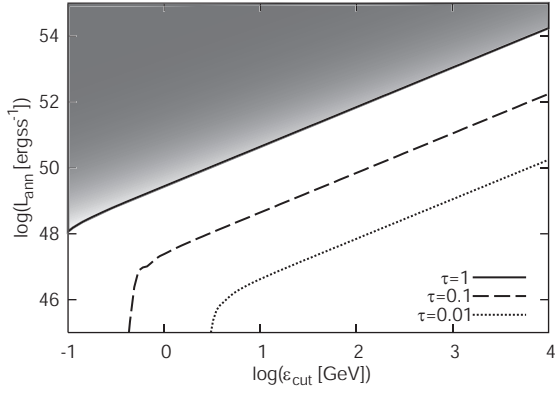}
\caption{
\footnotesize The same as Fig. 2, but for $L_p = {10}^{52}$ ergs
$\rm{s}^{-1}$. In this case, the fireball is not pair-dominated for 
sufficiently small $\varepsilon_{\rm{cut}}$, given fixed
$\tau$. When the fireball is baryon-dominated, we cannot expect
a pair-annihilation line. As pairs become dominant, $L_{\rm{ann}}$
increases sharply.  
}
\label{fig:spec3}
\end{figure}

\section{Cases for Limited Observations}\label{sec:limit}
\subsection{The Case of Non-detected Pair-Annihilation Lines}
\label{subsec:noline}
We can gain some information on the fireball even if a
pair-annihilation line is
not observed. The non-detection of pair-annihilation lines means 
\begin{eqnarray}
L_{\rm{ann}} \lesssim 
\varepsilon L_{\varepsilon}(\varepsilon_{\rm{ann}})=
L_{0}{ \left( \frac{\Gamma m_e
c^2}{(1+z) {\varepsilon}_{0}} \right) }^{2-\beta}
. 
\label{eq:Lann<}
\end{eqnarray}
If we can measure $L_{p}$, 
we can give upper limits on $\tau$, $2 n_{\pm}/n_p$ and $\tau_{\pm}$
by replacing $L_{\rm{ann}}$ with
$L_{0} (\Gamma m_e c^2/(1+z) \varepsilon_{0})^{2-\beta}$
in Eqs.~(\ref{eq:tau2}), (\ref{eq:npmnp}) and (\ref{eq:taupm2}).

Even when we cannot estimate $L_p$, the
inequalities~(\ref{eq:taupmtau}) where $L_p>0$ is used,
yield the looser constraint on the optical depth against pairs as
\begin{equation}
\tau_{\pm} \lesssim 
\left[\frac{16}{3 f_{\rm{cut}}}{\left( \frac{\Gamma m_e c^2}{(1+z) 
{\varepsilon}_{0}} \right)}^{2-\beta}
\right]^{1/2}.  
\label{eq:taupm<1}
\end{equation}
If the right hand side of the above inequality is smaller than unity,
i.e., ${\varepsilon}_{\rm{cut}} \gg 
%{[960(\beta-1)/11]}
{[16(\beta-1)/3 \xi(\beta)]}^{\frac{1}{\beta-1}} [\Gamma/(1+z)] m_e
c^2$, we can exclude the pair photospheric emission model. 
If we use $L_p \lesssim L_{\gamma}$ instead of $L_p>0$,
we obtain an upper limit on $\tau$ by replacing $L_p$ and $L_{\rm{ann}}$ 
with $L_{\gamma}$ and $L_{0} (\Gamma m_e c^2/(1+z) \varepsilon_{0})^{2-\beta}$,
respectively, in Eq.~(\ref{eq:tau2}). 

Note that we have implicitly assumed that $\Gamma$ is already determined 
by another means. At least we have $1 \leq \Gamma< (1+z) 
\varepsilon_{\rm{cut}}/m_e c^2$. $\Gamma$ can be estimated 
from $\tau_{\gamma \gamma}(\varepsilon_{\rm{cut}})=1$
in Eq.~(\ref{eq:taugg}) if we give the emission radius $r$. For
example, $r$ may be estimated from the frequently used relation
$r \approx 2 \Gamma^2 c {\delta t}_{\rm{decay}}/(1+z)$, where the
decay time of a pulse ${\delta t}_{\rm{decay}}$ is basically
determined by the angular spreading time scale \cite{Bar2,Lit1}. 
The possible thermal emission component may be also useful to estimate
$\Gamma$ \cite{Pee4}.

\subsection{The Case of Non-detected Cutoff Energy}\label{sec:nocut} 
Because of the limited sensitivity of the detector, 
the observed maximum energy ${\varepsilon}_{\rm{max}}$ may be 
smaller than the true cutoff energy ${\varepsilon}_{\rm{cut}}$. 
As seen in Eq. (\ref{eq:fdef}), $f(\varepsilon,\Gamma)$ increases 
with $\varepsilon$ for $\varepsilon \lesssim \varepsilon_{\rm{cut}}$,  
as long as the cutoff energy is determined by the pair-creation 
process. (More precisely, $\tau_{\gamma \gamma} (\varepsilon)$, hence 
$f(\varepsilon,\Gamma)$ typically reaches almost the maximum value 
around $\varepsilon \sim \tilde{\varepsilon}_{\rm{peak}}$ for 
the low energy spectral index $\alpha \lesssim 1$, where 
$\varepsilon_{\rm{peak}}$ is the peak energy. On the other hand, 
$\tau_{\gamma \gamma} (\varepsilon)$ always increases with 
$\varepsilon$ for $\alpha \gtrsim 1$.) 
Then, we have
\begin{equation}
f({\varepsilon}_{\rm{cut}},\Gamma) \gtrsim 
f({\varepsilon}_{\rm{max}},\Gamma).
\label{eq:fcut>}
\end{equation}
If we can measure $L_{p}$, we can give upper limits on $\tau$ and $\tau_{\pm}$
as well as an lower limit on $2 n_{\pm}/n_p$ by replacing $f_{\rm{cut}}$ with
$f_{\max} \equiv f({\varepsilon}_{\rm{max}},\Gamma)$
in Eqs.~(\ref{eq:tau2}), (\ref{eq:npmnp}) and (\ref{eq:taupm2}). 

Without knowing $L_p$, the inequalities~(\ref{eq:taupmtau}), where
$L_p>0$ is used, yield the looser upper limit on $\tau_{\pm}$ as
\begin{equation}
\tau_{\pm} \lesssim 
\left(\frac{16}{3} \frac{L_{\rm{ann}}}{L_{0} f_{\rm{max}}}
\right)^{1/2}.
\label{eq:taupm<2}
\end{equation} 
If the right hand side of the above inequality is less than unity, i.e.,
${\varepsilon}_{\rm{max}} \gg 
%{[960(\beta-1)/11]}
{[16(\beta-1)/3 \xi(\beta)]}^{\frac{1}{\beta-1}} [\Gamma  m_e c^2/(1+z)] 
L_{\rm{ann}} L_{0}^{-1} {[\Gamma m_e
c^2/(1+z) {\varepsilon}_{0}]}^{2-\beta}$, 
the pair photospheric emission model is ruled out.
If we use $L_p \lesssim L_{\gamma}$ instead of $L_p>0$,
we obtain an upper limit on $\tau$ 
as well as a lower limit on $2 n_{\pm}/n_p$
by replacing $L_p$ and $f_{\rm{cut}}$ with
$L_{\gamma}$ and $f_{\max}$, respectively,
in Eqs.~(\ref{eq:tau2}) and (\ref{eq:npmnp}).

The above arguments in this subsection can be applied even when the
fireball is completely thin, i.e., the cutoff energy due to 
the pair-creationprocess in the source does not exist.
If we know that this is the case from other means, we can replace 
$f_{\rm{max}} \equiv f(\varepsilon_{\rm{max}}, \Gamma)$ with
$f(\tilde{\varepsilon}_{\rm{peak}}, \Gamma)$ 
in the inequality~(\ref{eq:fcut>}) for $\alpha \lesssim 1$.

\section{Implications}\label{sec:implications}
In this paper, we have clarified that pair-signatures provide 
useful information on the fireball in GRBs only with observable quantities.  
The strategy for acquiring the physical quantities is 
summarized in the caption of Fig. 1. 

\subsection{Examination of $r$ and $\Gamma$}
%Hence, the emission radius $r$ may be regarded as a generic parameter. 
The determination of emission radii $r$ is important not only for
specifying the model of prompt emissions but also for various
model-predictions (e.g., neutrino production in the internal shock
model is senstive to emission radii
$r$ \citep[e.g., ][]{Mur1,Mur3}). After this work on pair-signatures by us, 
Gupta \& Zhang (2007) recently focused on this issue of the unknown 
emission radius. They re-expressed the cutoff energy as a function of 
$r$ and $\Gamma$. By using Eqs. (\ref{eq:taugg}) and (\ref{eq:ng}), 
we can see that the emission radius $r$ is obtained from 
observationally determined $\varepsilon_{\rm{cut}}$, $\varepsilon_0$ 
and the radiation energy of a subshell at $\varepsilon_0$, 
$E_0 \sim L_0 {\delta t}_{\rm{rise}}/(1+z)$, if we know $\Gamma$ 
by other means. Here, $\delta t_{\rm rise}$ is the rise time of a
pulse, which is basically determined by the comoving width of the
subshell $\Delta' \approx \Gamma c \delta t_{\rm rise}/(1+z)$.
Eq. (1) is one of the ways to determine $\Gamma$. Other means 
(e.g., by using the photospheric emission component \cite[]{Pee4}) 
are also useful. 

On the other hand, the emission radius can be also estimated via the
relation $r \approx 2 \Gamma^2 c {\delta t}_{\rm{decay}}/(1+z)$, as
noted in \S~\ref{subsec:noline}. Once this relation is validated, we 
can compare the emission radius estimated 
by using this relation with that determined from 
$\varepsilon_{\rm{cut}}$. 
In other words, we can test whether 
$\Gamma$ determined by Eq. (\ref{eq:tau1}) and $r \approx 2 \Gamma^2 
c {\delta t}_{\rm{decay}}/(1+z)$ is consistent with $\Gamma$ estimated 
from Eq. (1) and other means or not. Because the derived
$\Gamma$ should be consistent if the emission radius is the same,
they will be useful as another closure relation (see
\S~\ref{sec:discussion}).    
Note that we have not so far assumed the relation $r
\approx 2 \Gamma \Delta' \approx 2 \Gamma^2 c {\delta
t}_{\rm{rise}}/(1+z)$, which is expected in the internal shock model 
%we could relate the width to the variabiity time scale as 
%$\Delta^{\prime} \approx c t_{\rm{dyn}}/(1+z) = \Gamma c 
%{\delta t}_{\rm{var}}/(1+z)$, 
but may not be true. In fact, models other than the 
internal shock model do not always predict 
$r \approx 2 \Gamma \Delta^{\prime}$, but can lead to 
$r \gg 2 \Gamma \Delta^{\prime}$. 

\begin{figure}[bt]
\plotone{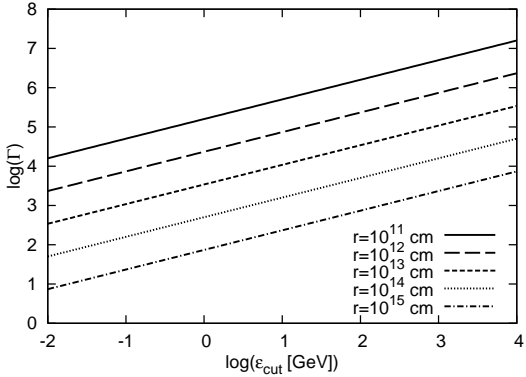}
\caption{
\footnotesize The relation between the cutoff energy
$\varepsilon_{\rm{cut}}$ and bulk Lorentz factor $\Gamma$ for 
given collision radii $r$. A simple power-law photon spectrum is
assumed. Used parameters are $E_0={10}^{51}$ 
ergs, $\varepsilon _0 = {10}^{2.5}$ keV and $z=0.1$. 
Note that $\varepsilon_{\rm{cut}} \lesssim 
\tilde{\varepsilon}_{\rm{peak}}$ is also assumed implicitly. 
}
\label{fig:spec4}
\end{figure}

\subsection{Test of the Pair Photospheric Emission Model}
As already noted, pair-signatures are especially useful to test the 
pair photospheric emission model, where the prompt emission comes from 
$r_{\rm{ph}} \approx r_{\rm{ph},\pm}$. 
We can measure $\tau_{\pm}$ by Eq.~(\ref{eq:taupm2}) with $L_p$,
and an upper limit on $\tau_{\pm}$ by the inequalities~(\ref{eq:taupmtau})
without $L_p$. If we can observe either the 
pair-annihilation line or the cutoff energy due to pair-creation, 
an upper limit on $\tau_{\pm}$ is obtained by Eq.~(\ref{eq:taupm2}) 
with the inequalities~(\ref{eq:Lann<}) or (\ref{eq:fcut>}) 
for known $L_p$, and by the inequality~(\ref{eq:taupm<1}) or 
(\ref{eq:taupm<2}) for unknown $L_p$. 
When the fireball is pair-dominated, i.e., the inequality
(\ref{eq:pairdom}) is satisfied, we have $\tau \approx
\tau_{\pm}$. In addition, under the photospheric emission model, 
we expect that high-energy gamma-rays are
produced by the dissipation around the photosphere (which may occur at
the sub-photosphere) and emerge from the emission region at
$r \sim r_{\rm{ph}}$. Therefore, the pair photospheric emission 
model predicts $\tau \approx \tau_{\pm} \sim 1$ 
in Eqs.~(\ref{eq:taupm}) or (\ref{eq:tau2}). 
 
When the fireball is pair-dominated, the photospheric radius where 
$\tau \approx \tau_{\pm} \simeq 1$ can be expressed as
\begin{equation} 
r_{\mr{ph}}\approx r_{\rm{ph},\pm} \simeq 
\frac{f_{\rm{cut}} L_{0} \sigma_T}{4 \pi m_e c^3 \Gamma^3 q},
\label{eq:rpm}
\end{equation}
where $q \equiv r_{\rm{ph}}/\Gamma \Delta^{\prime}$ which is expected
to be an order-of-unity factor in the internal shock model. 
Eq. (\ref{eq:rpm}) is essentially the same
equation as that shown in Rees \& Meszaros (2005). 
Note that the relation $r \approx 2 \Gamma \Delta'$ 
expected in the internal shock model leads to $q=2$.

When $L_{\rm{ann}} \sim  \varepsilon L_{\varepsilon}(\varepsilon_{\rm{ann}})
=L_{0} {[\Gamma m_e c^2/(1+z) 
{\varepsilon}_{0}]}^{2-\beta}$, the pair photospheric emission model,  
under which we expect $\tau \approx \tau_{\pm} \sim 1$, predicts the unique
relation between ${\varepsilon}_{\rm{cut}}$ and ${\varepsilon}_{\rm{ann}}$.
Eq. (\ref{eq:taupm}) yields (after the integration over $\varepsilon_{\gamma}$
in Eq.~(\ref{eq:fdef}) which is the expression of $f_{\rm{cut}}$),
\begin{equation}
\frac{{\varepsilon}_{\rm{cut}}}{{\varepsilon}_{\rm{ann}}} 
\sim {\left[ \frac{(\beta-1)}{\tau_{\pm}^2}
\frac{16}{3 \xi (\beta)} \right]}^{\frac{1}{\beta-1}}
\sim {\left[ (\beta-1)
\frac{16}{3 \xi (\beta)} \right]}^{\frac{1}{\beta-1}}.
\label{eq:ecuteann}
\end{equation}
If pairs are created by the underlying continuum photons,
the pair-annihilation line cannot exceed the continuum emission
much prominently, i.e., $L_{\rm{ann}} \sim L_{0} {[\Gamma m_e c^2/(1+z) 
{\varepsilon}_{0}]}^{2-\beta}$ \cite{Iok1,Pee1,Pee3}.
Therefore, the relation (\ref{eq:ecuteann}) could be satisfied for
many bursts under the pair photospheric emission model. 
Superposing low-quality spectra of many events
by adjusting either ${\varepsilon}_{\rm{ann}}$ or 
${\varepsilon}_{\rm{cut}}$ could help to find the other feature
in this model.

\section{Discussion}\label{sec:discussion}
Although the pair-signatures give us useful information, we have to be
careful because there are some uncertainties in obtained quantities
% in 
and we have put several assumptions in deriving equations and
inequalities shown in this paper.

First, we have assumed that all the photons come from the same
emission region. However, this might not be true.
Although we assume the same emission radius for pair-signatures as a
first step consideration, actual emissions may not come from the same 
emission radius. For example, let us consider cases where high-energy
gamma-rays come from two different emission radii $r_1$ and $r_2$
($r_1 < r_2$). There will be three possibilities; (A) 
Case where the observed pair-annihilation line comes from $r_1$,
while the pair-creation cutoff coming from $r_2$,
$\varepsilon_{\rm{cut},2}$ is higher than that from $r_1$, 
$\varepsilon_{\rm{cut},1}$; (B) Case where 
the observed pair-annihilation line comes from $r_2$, while the pair-creation 
cutoff coming from $r_1$, $\varepsilon_{\rm{cut},1}$ is higher than 
that from $r_2$, $\varepsilon_{\rm{cut},2}$; (C) Case where 
both of the observed pair-annihilation line and (higher) pair-creation
cutoff come from $r_1$ or $r_2$.

(A-1) Case (A) where the underlying continuum dominantly comes from
$r_1$ at $\varepsilon \lesssim \varepsilon_{\rm{cut},1}$;
In this case, we will ideally see $\varepsilon_{\rm{cut},1}$ coming
from $r_1$ as well as $\varepsilon_{\rm{cut},2}$ which is the higher cutoff. 
Because we have higher $\tau_{\gamma \gamma}$ at smaller $r$ (and/or
for larger $E_0$), with the given $\Gamma$, the former cutoff could be
naturally lower than the latter. If we could see two
$\varepsilon_{\rm{cut}}$ in the photon 
spectrum (i.e., one is due to $\varepsilon_{\rm{cut},1}$, while the other 
is $\varepsilon_{\rm{cut},2}$ which is higher), our recipes would be 
applied to the line and the lower cutoff $\varepsilon_{\rm{cut},1}$. 
(A-2) Case (A) where the underlying continuum dominantly comes from
$r_2$ at $\varepsilon \lesssim \varepsilon_{\rm{cut},2}$; 
%In this case, it would be difficult to
%judge whether emission comes from the same emission radius or not. 
In this case, and if the outflow has similar $\Gamma$ at the two emission
radii, we would expect that time-resolved detailed observations could 
separate different emission radii. It is because, if we can use 
$r \approx 2 \Gamma^2 c {\delta t}_{\rm{decay}}/(1+z)$, 
the larger emission radius $r$ leads to
the longer ${\delta t}_{\rm{decay}}$. 
On the other hand, if the outflow has different $\Gamma$ at the two emission
radii, it is useful 
%we could expect that 
to determine $\Gamma$ independently in various ways.
% will be useful.
%In addition, we would also need to
%evaluate the Lorentz factor for each pair-signature as precisely as possible. 
Although it may be observationally difficult, not only the estimation 
by using Eq. (1) but also other means for estimation (e.g., by using
the photospheric emission component and/or the relation $r \approx 
2 \Gamma^2 c {\delta t}_{\rm{decay}}/(1+z)$), would enable us 
to evaluate $\Gamma$. If emissions come from the same emission
radius, we expect that all of the $\Gamma$ we obtain should be
consistent. 
(B-1) Case (B) where the underlying continuum dominantly comes from
$r_2$ at $\varepsilon \lesssim \varepsilon_{\rm{cut},2}$; Since the 
pair-creation cutoff from $r_1$ is higher, not completely masked by the
underlying continuum from $r_2$, we can see $\varepsilon_{\rm{cut},2}$
below $\varepsilon_{\rm{cut},1}$ as in the case (A-1). 
We may apply our recipes
to the line and the lower cutoff $\varepsilon_{\rm{cut},2}$.
(B-2) Case (B) where the underlying continuum dominantly comes from
$r_1$ at $\varepsilon \lesssim \varepsilon_{\rm{cut},1}$;
In this case, the higher cutoff $\varepsilon_{\rm{cut},1}$ comes from 
the inner radius $r_1$, while $\varepsilon_{\rm{cut},2}$ is masked and
the observed pair-annihilation line is generated at the larger radius
$r_2$. It would require, typically, that the Lorentz factor at the
outer emission radius $r_2$ is smaller than that at the inner emission
radius $r_1$, because the prominent pair-annihilation line and the lower 
$\varepsilon_{\rm{cut},2}$ would mean copious pairs and photons at $r_2$. 
Therefore, evaluation of Lorentz factors by several means 
would be important.
(C-1) Case (C) where both pair-signatures come from $r_1$ while the 
underlying continuum dominantly comes from $r_2$ 
at $\varepsilon \lesssim \varepsilon_{\rm{cut},2}$; In this case, we
can see $\varepsilon_{\rm{cut},1}$ above $\varepsilon_{\rm{cut},2}$ in
principle. Hence, our recipes can be applied to the line and higher cutoff,
while if we use the lower cutoff, we could obtain the Lorentz factor 
that is inconsistent with other estimations. (C-2) Case (C) where
both pair-signatures come from $r_2$ while the 
underlying continuum dominantly comes from $r_1$ 
at $\varepsilon \lesssim \varepsilon_{\rm{cut},1}$; Similarly to the
case (C-1), we could two $\varepsilon_{\rm{cut}}$ ideally.
We can apply our recipes to the higher cutoff,
and then obtain the Lorentz factor 
that is consistent with the estimation by Eq.~(1).

Therefore, what we have to do is that we apply our recipes to 
the time-resolved spectra, if possible, and 
then compare the Lorentz factors obtained by several means in order to
check the consistency. When we have two or more cutoffs, we may select
the cutoff that provides the consistent Lorentz factor.
Once we can see that emissions come from the same radius, our recipes
described 
in this paper can be used in order to obtain information on
the fireball of GRBs 

Second, we have assumed that sufficiently relativistic electrons cool 
down rapidly, $t_{\rm{cool}} \ll t_{\rm{ann}}$, which is
expected in many models \cite{Iok1,Pee1,Pee2}. However, 
the pair-annihilation line might come
from relativistic pairs. 
For example, in the slow dissipation
scenario \cite{Ghi1}, e.g., as might be expected from magnetic
reconnection, the typical electron Lorentz factor at the
end of the dynamical time $\gamma_{\rm{cool}}$ could be larger
than unity \cite{Pee3}. If $\gamma_{\rm{cool}} > 1$, we should use 
${\varepsilon}_{\rm{ann}} \sim \Gamma \gamma_{\rm{cool}} m_e c^2 /(1+z)$ 
instead of Eq. (1), and the expression of $L_{\rm{ann}}$ should 
also be modified 
(where $L_{\rm{ann}}$ is suppressed for $\gamma_{\rm{cool}} 
\beta_{\rm{cool}}^2 / (1+ \beta_{\rm{cool}}^2) \gtrsim 1$) \cite{Sve1}.
In such a case, it becomes more difficult to observe the 
pair-annihilation line since the width of the pair-annihilation line 
is broadened by more than order-of-unity in energy due to the broad 
distributions of relativistic pairs, although we may check this 
observationally \cite{Sve1}. 
If we can specify the distributions of electrons and positrons properly 
(e.g., thermal distributions), we could evaluate $L_{\rm{ann}}$,
$\Gamma$ and the shape of the pair-annihilation line with elaborate
observational results in the future. But in the case where the distributions of
electrons and positrons are unknown, they are model-dependent as 
demonstrated in Pe'er et al. (2005,2006), which would cause possible 
ambiguities for our recipes.   

Third, we have also assumed that the cutoff energy
$\varepsilon_{\rm{cut}}$ is determined by attenuation
via $\gamma \gamma \rightarrow e^{+} e^{-}$ in the source. 
However, the attenuation due to interaction with cosmic infrared
background photons should be also taken into account when 
$\varepsilon_{\rm{cut}}$ is sufficiently high. This cosmic attenuation effect
can make it difficult to determine the cutoff energy at the source, 
$\varepsilon_{\rm{cut}}$. The observed maximum energy might also represent 
the maximum energy of accelerated electrons.
In order to evaluate $\varepsilon_{\rm{cut}}$ properly,
the careful analyses will be needed. 
The secondary delayed emission may be also useful \cite{Mur2}.
Note that we can apply the recipe in (\S~\ref{sec:nocut}) even 
without the true $\varepsilon_{\rm{cut}}$.

Pair-signatures may be detected by the future-coming GLAST satellite.
However, the detection of pair-annihilation lines may be difficult due
to line-broading, as discussed in (\S~\ref{sec:test}). Lines are observed 
as bumps, so that evaluated $\tau$ and $L_{\pm}$ will have uncertainties 
by a factor, due to observational difficulties in precise determination 
of $L_{\rm{ann}}$ and $\Gamma$.
In addition, it is hopeful to apply our recipes to single pulses. 
Some GRBs can be regarded as single pulse events. 
For example, some bright bursts such as BATSE trigger numbers 647 and
999 exhibited relatively smooth, long, single pulses, which were
separated well from other pulses. For such single pulses, 
we may expect emissions from the approximately same emission radius, although 
the spectrum also showed the time-dependent evolution. Note that, our 
recipes could be applied to flares, where wider and smoother pulses are seen
\cite{Bur05,ioka05}. 
A flare may come from the approximately same emission radius. However, the 
detection of pair-annihilation lines will be more difficult
observationally because pair-annihilation lines from flares are
typically expected at $\sim 10$ MeV if the Lorentz factor of
flare-outflows is $\sim 10$. Furtheremore, emissions from 
flares will be contaminated by afterglow components.

The height of the pair-annihilation line may be 
comparable to the underlying continuum emission. Therefore, 
we have to collect 
sufficiently many photons to identify the pair-annihilation line. 
For example, if the height of the pair-annihilation line is larger 
than the underlying continuum by a factor 
$\sim 2$, we need to collect $\sim 20$ 
photons for the 3 $\sigma$ detection at $\sim {\varepsilon}_{\rm{ann}}$. 
When the spectrum of the prompt emission is expressed
by a power-law extending to sufficiently high energies, GLAST/LAT is expected
to find $\sim 70$ GRBs per year under the criterion that $>10$ photons 
per bursts are collected for the energy threshold $30$ MeV
\cite{Omo1}. It suggests that, if a significant fraction of GRBs 
accompanies pair-annihilation lines, we expect good opportunities to
see them.
 
We also have to note that
there may be some uncertainties in determining
${\varepsilon}_{\rm{cut}}$. Opacity
skin effects can sometimes render the exponential attenuation 
$\exp(-\tau_{\gamma \gamma})$ a poor descriptor of attenuation with 
$1/(1+\tau_{\gamma \gamma})$, which leads to broken power-laws rather
than exponential turnovers \cite{Bar1,Bar2}. 
We expect that such ambiguities could be solved by
observing the maximum energy for a lot of events. 
     
%% If you wish to include an acknowledgments section in your paper,
%% separate it off from the body of the text using the \acknowledgments
%% command.
%% Included in this acknowledgments section are examples of the
%% AASTeX hypertext markup commands. Use \url without the optional [HREF]
%% argument when you want to print the url directly in the text. Otherwise,
%% use either \url or \anchor, with the HREF as the first argument and the
%% text to be printed in the second.

\acknowledgments
We thank S. Nagataki, T. Nakamura and K. Toma for helpful comments.
This work is supported in part by the Grant-in-Aid for the 21st Century COE
"Center for Diversity and Universality in Physics" from the Ministry
of Education, Culture, Sports, Science and Technology (MEXT) of Japan
and by the Grant-in-Aid from the 
Ministry of Education, Culture, Sports, Science and Technology
(MEXT) of Japan, No.18740147, No.19047004 (K.I.).
K.M. is supported by Grant-in-Aid for JSPS Fellows.

\clearpage

%% Use the figure environment and \plotone or \plottwo to include
%% figures and captions in your electronic submission.
%% To embed the sample graphics in
%% the file, uncomment the \plotone, \plottwo, and
%% \includegraphics commands
%%
%% If you need a layout that cannot be achieved with \plotone or
%% \plottwo, you can invoke the graphicx package directly with the
%% \includegraphics command or use \plotfiddle. For more information,
%% please see the tutorial on "Using Electronic Art with AASTeX" in the
%% documentation section at the AASTeX Web site,
%% http://www.journals.uchicago.edu/AAS/AASTeX.
%%
%% The examples below also include sample markup for submission of
%% supplemental electronic materials. As always, be sure to check
%% the instructions to authors for the journal you are submitting to
%% for specific submissions guidelines as they vary from
%% journal to journal.

%% This example uses \plotone to include an EPS file scaled to
%% 80% of its natural size with \epsscale. Its caption
%% has been written to indicate that additional figure parts will be
%% available in the electronic journal.

%% If you are not including electonic art with your submission, you may
%% mark up your captions using the \figcaption command. See the
%% User Guide for details.
%%
%% No more than seven \figcaption commands are allowed per page,
%% so if you have more than seven captions, insert a \clearpage
%% after every seventh one.

%% The following command ends your manuscript. LaTeX will ignore any text
%% that appears after it.

\end{document}